\documentclass[prb,twocolumn,showpacs]{revtex4}     
\usepackage{epsfig,amsmath} 

\begin{document}

\title{Compressibility Instability of Interacting Electrons in Bilayer Graphene} 

\author{Xin-Zhong Yan$^{1}$ and C. S. Ting$^2$}
\affiliation{$^{1}$Institute of Physics, Chinese Academy of Sciences, P.O. Box 603, 
Beijing 100190, China\\
$^{2}$Texas Center for Superconductivity, University of Houston, Houston, Texas 77204, USA}
 
\date{\today}
 
\begin{abstract}
Using the self-consistent Hartree-Fock approximation, we study the compressibility instability of the interacting electrons in bilayer graphene. The chemical potential and the compressibility of the electrons can be significantly altered by an energy gap (tunable by external gate voltages) between the valence and conduction bands. For zero gap case, we show that the homogeneous system is stable. When the gap is finite, the compressibility of the electron system becomes negative at low carrier doping concentrations and low temperature. The phase diagram distinguishing the stable and unstable regions of a typically gapped system in terms of temperature and doping is also presented.
\end{abstract}


\maketitle

Bilayer graphene has attracted considerable attention because of its promised application in electronic devices. \cite{Ohta, Henriksen, Young, Martin, McCann1, Min, McCann2, Koshino, Nilsson1, Barlas1, Nilsson2, Wang, Hwang, Barlas2, Borghi1, Nandkishore, Borghi2, Kusminskiy1, Borghi3} In contrast to the Dirac fermions in monolayer graphene, the energy bands of the free electrons in bilayer graphene are hyperbolic and gapless between the valence and conduction bands.\cite{McCann1} Most importantly, an energy gap opening between the valence and conduction bands can be generated and controlled by external gate voltages. At low carrier doping, because the energy-momentum dispersion relation around the Fermi level is relatively flat, the Coulomb effect is expected to be significant in the interacting electron system of bilayer graphene. The Coulomb effect has been studied by a number of theoretical works using the Hartree-Fock (HF) and random-phase approximations.\cite{Nilsson2, Wang, Hwang, Barlas2, Borghi1, Nandkishore, Borghi2, Kusminskiy1, Borghi3} One of the thermodynamic quantities directly reflecting the Coulomb effect is the electronic compressibility. Recently, this quantity has been measured by experiments.\cite{Henriksen,Young,Martin} For timely coordinating with the experimental measurements, it is necessary to theoretically study the combined effect of Coulomb interactions and the gap opening in the compressibility.

In this work, using the self-consistent HF (SCHF) approximation,\cite{Borghi3} we investigate the energy bands, the chemical potential and its derivative with respect to the carrier density for the interacting electrons in bilayer graphene. With these results, the compressibility instabilities for gapped and ungapped electron systems are examined. We show that the compressibility is always positive for zero gap case, while for the gapped system the compressibility becomes negative at low doping and low temperature that implies the system would become inhomogeneous. The phase diagram distinguishing the stable and unstable regions in the temperature-density plane is presented for a typically gapped homogeneous electron system in bilayer graphene. SCHF is a conserving approximation satisfying microscopic conservation laws.\cite{Baym} The conserving conditions are crucial in the many particle systems. 
 
\begin{figure} 
\centerline{\epsfig{file=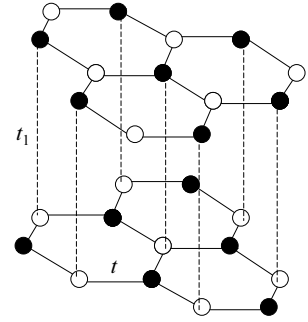,width=4.5 cm}}
\caption{Structure of Bernal stacking bilayer graphene. The energies of intra- and interlayer electron hopping between the A (white) and B (black) atoms are given by $t$ and $t_1$, respectively. } 
\end{figure} 

The atomic structure of bilayer graphene is shown in Fig. 1. The two sublattices in each layer are denoted by A (white) and B (black) atoms, respectively. The interlayer distance is $c = 3.34$ \AA~ $\approx 1.4 a$ where $a \approx 2.4$ \AA~is the lattice constant of monolayer graphene. The energy of electron hopping between the nearest-neighbor (nn) carbon atoms in each layer is $t \approx 2.82$ eV,\cite{Bostwick} while the interlayer nn hopping is $t_1 \approx 0.39$ eV.\cite{Misu} 

The Hamiltonian describing the electrons is given by
\begin{eqnarray}
H = \sum_{p}\psi^{\dagger}_{p\sigma}h^0\psi_{p\sigma}+\frac{1}{2}\sum_{ij}n_iv_{ij}n_j \label {hm}
\end{eqnarray}
where $\psi^{\dagger}_{p\sigma} = (c^{\dagger}_{p,a,1},c^{\dagger}_{p,b,1},c^{\dagger}_{p,a,2},c^{\dagger}_{p,b,2})$ with $p = (k,v,\sigma)$ standing for the momentum $k$, the valley $v$ and spin $\sigma$, and $c^{\dagger}_{p,a,\ell}$ creating an electron in $p$ state at $a$ lattice of the $\ell$ layer (with the top layer labeled as 1), $n_i$ is the electron density at the site $i$, and $v_{ij}$ is the Coulomb interaction between electrons at sites $i$ and $j$. Here, for the low energy electrons in bilayer graphene with interlayer hopping, the matrix $h^0$ is an extension \cite{McCann1,Koshino} of the Dirac fermions in monolayer graphene.\cite{Wallace,Yan} By denoting $k_x+ ik_y = k\exp(i\phi)$, $h^0$ can be written as $h^0 = T(\phi)h(k)T^{\dagger}(\phi)$ with $T(\phi)$ = Diag(1,$e^{i\phi},e^{i\phi},e^{i2\phi}$) and 
\begin{eqnarray}
h(k) = \begin{pmatrix}
\Delta & k &0 &0 \\
k& \Delta &t_1 &0 \\
0 &t_1 &-\Delta &k \\
0 &0 &k & -\Delta \\
\end{pmatrix} 
\end{eqnarray}
in units of $E_0 \equiv \sqrt{3}t/2 = a  = 1$. Here $\Delta$ is the energy gap parameter, which is a consequence of the potential difference between the top and back gates (attached to the top and lower layers, respectively). The form of $h^0$ reminds us to take the transform $\psi_{p\sigma} = T(\phi)\Phi_{p\sigma}$ and then work with the scalar momentum $k$. We here take the cutoff of $k$ as $k_c = 1$. Under the SCHF approximation to the Coulomb interactions, one obtains an effective Hamiltonian 
\begin{eqnarray}
H_{\rm eff} = \sum_{p}\Phi^{\dagger}_{p}h_{\rm eff}(k)\Phi_{p} 
\end{eqnarray}
with $h_{\rm eff}(k) = h(k) + \Sigma(k)$ and 
\begin{eqnarray}
\Sigma_{\mu\nu}(k) &=& -\frac{1}{V}\sum_{k'}v_{\mu\nu}(|\vec k-\vec k'|)\cos\theta_{\mu\nu}\langle \Phi^{\dagger}_{p'\nu}\Phi_{p'\mu}\rangle, \label{slf} \\
(\theta_{\mu\nu})&=& \begin{pmatrix}
0& \alpha &\alpha &2\alpha \\
\alpha& 0 &0 &\alpha \\
\alpha &0 &0 &\alpha \\
2\alpha &\alpha &\alpha & 0 \\
\end{pmatrix}
\end{eqnarray}
where $v_{\mu\nu}(|\vec k-\vec k'|)$ is the Coulomb interaction, $\alpha$ is the angle between $\vec k$ and $\vec k'$, and $\Phi_{p'\mu}$ [with $p' = (k',v,\sigma)$] is the $\mu$th component of $\Phi_{p'}$. For electrons in the same layer, $v_{\mu\nu}(q) = 2\pi e^2/\epsilon q$ with $\epsilon = 4$. For interlayer electron interactions, $v_{\mu\nu}(q) = 2\pi e^2\exp(-cq)/\epsilon q$. The self-energy $\Sigma(k)$ is diagrammatically shown in Fig. 2 (a) in terms of the Green's function. In the numerical calculations, special care needs to be paid to a logarithmic singularity stemming from the azimuthally integral of $v_{\mu\nu}(|\vec k-\vec k'|)\cos\theta_{\mu\nu}$. 

\begin{figure} 
\centerline{\epsfig{file=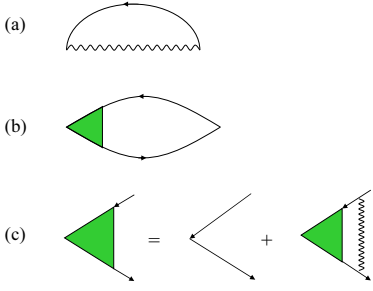,width=5.5 cm}}
\caption{(color online) Diagrammatic expressions for (a) the self-energy $\Sigma(k)$, (b) the irreducible density-density response function, and (c) the vertex correction. The solid line with an arrow is the Green's function and the wavy line represents the Coulomb interaction.} 
\end{figure} 

$H_{\rm eff}$ can be diagonalized by using the transformation 
\begin{eqnarray}
\Phi_{p\mu} = \sum_{\nu}U_{\mu\nu}(k)V_{p\nu} 
\end{eqnarray}
where $U(k)$ is an unitary matrix and $V_{p\nu}$ is the eigen operator of a quasiparticle with energy $E_{\nu}(k)$. The chemical potential $\mu$ is determined by the doped electron density
\begin{eqnarray}
n = \frac{4}{V}\sum_{k}\{\sum_{\nu}f[E_{\nu}(k)] -2\}
\end{eqnarray}
where the factor 4 comes from the spin and valley degeneracy, and $f(E)= 1/\{\exp[(E-\mu)/T]+1\}$ is the Fermi distribution function. The diagonalization of $H_{\rm eff}$ and determination of $\Sigma_{\mu\nu}$ and $\mu$ are carried out by iteration until the self-consistency is achieved.

\begin{figure} 
\centerline{\epsfig{file=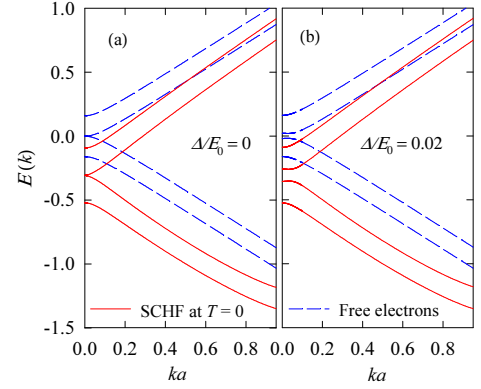,width=6.5 cm}}
\caption{(color online) Energy band structure of free electrons (dashed lines) and the interacting electrons under the self-consistent Hartree-Fock approximation at $T = 0$ (solid lines) for (a) $\Delta/E_0 = 0$ and $\delta = 0$ and (b) $\Delta/E_0 = 0.02$ and $\delta = 10^{-5}$. } 
\end{figure} 

In Fig. 3, we show the energy band structure of the electrons at temperature $T = 0$ for (a) $\Delta = 0$ and electron doping concentration $\delta \equiv \sqrt{3}a^2n/8$ (doped electrons per carbon atom) = 0 and (b) $\Delta/E_0 = 0.02$ (the typical magnitude in experiments \cite{Henriksen,Young}) and $\delta = 10^{-5}$. The solid lines are the energies of the electrons renormalized by the Coulomb interactions. The four energy bands of free electrons denoted by the dashed lines are given by $E^0(k) = \pm [\Delta^2+t^2_1/2+k^2 \pm\sqrt{t^4_1/4 + (t^2_1+4\Delta^2)k^2}]^{1/2}$. The upper and lower bands of the free electrons are symmetric about zero energy. For the interacting electrons, the energies shift downwards because of the self-energy contributions, and the upper and lower bands are not symmetric anymore. For the zero-gap case, the zero gap is not changed by Coulomb interactions. Notice that the contribution from the two lower bands to the self-energy is included here. If it is a constant for any doping and temperature, it can be subtracted from the beginning. However, in the SCHF, the valence band itself and thereby its contribution to the self-energy vary with doping and temperature. Especially, in the zero gap case, this variation is significant. The contribution from all the two lower bands cannot be considered as a constant and subtracted from the beginning. 

\begin{figure} 
\centerline{\epsfig{file=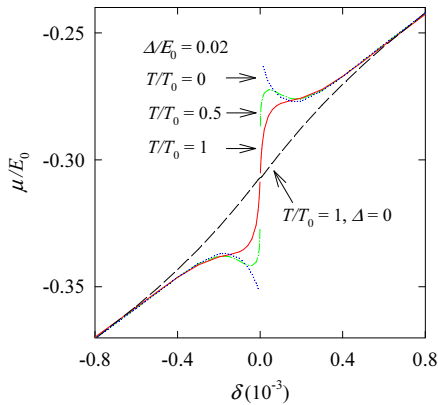,width=6.1 cm}}
\caption{(color online) Chemical potential $\mu$ as function of doping concentration $\delta$ for $\Delta/E_0 = 0$ at $T/T_0 = 1$ (black dashed line) and $\Delta/E_0 = 0.02$ at $T/T_0 = 0$ (blue dotted line), $T/T_0 = 0.5$ (green dash-dot line) and $T/T_0 = 1$ (red solid line). } 
\end{figure} 

Figure 4 shows the chemical potential $\mu$ as a function of $\delta$ for the system with $\Delta/E_0 =0.02$ at several temperatures and that of the zero-gap system at $T/T_0 = 1$ with $T_0 \equiv 0.01E_0 \approx 283$ K (close to the room temperature). For the gapped system, the low doping behavior of $\mu$ is delicate. At $T = 0$, $\mu$ varies discontinuously at $\delta = 0$. This discontinuity stems from the energy gap between the valence and conductions bands. From both sides of the carrier doping close to $\delta = 0$, $\mu$ is not a monotonic function of $\delta$. In contrast to this zero $T$ behavior, at finite $T$, $\mu$ increases dramatically at small doping. It is well known that $\mu$ at finite $T$ is lower than that at $T = 0$ for a one-band electron system with parabolic energy-momentum dispersion relation because the particles occupy energy levels above the Fermi energy with a lowered chemical potential. In the present gapped case, the conduction band is the effective one band system. At smaller electron doping, the Fermi energy is lower, the temperature effect is much more pronounced. The situation for the zero-gap system is different. The temperature effect in $\mu$ of the zero-gap system is not so notable as in the gapped system. For $\Delta = 0$, the electrons in the valence band can be thermally excited to high levels in the conduction band without significant change in the chemical potential because the density of states is nearly symmetric about the touch point of the valence and conduction bands. The chemical potential for $\Delta = 0$ at $T/T_0 = 1$ is almost the same as at $T = 0$. 

The behavior of the chemical potential is closely related to the compressibility $\kappa$ and the spin susceptibility $\chi$ of the doped carrier system. They are defined as
\begin{equation}
\kappa = \frac{1}{n^2}(\frac{\partial n}{\partial\mu})_T   \label{cmp}			
\end{equation}
and
\begin{equation}
\chi = \frac{\hbar\mu_B}{2}(\frac{\partial n}{\partial\mu})_T   \label{ss}				
\end{equation}
with $\mu_B$ as the Bohr magneton. The common factor $(\partial n/\partial\mu)_T$ can be calculated by performing the derivative with the obtained result as shown in Fig. 4. On the other hand, this factor is actually the irreducible density-density response function $\chi_{nn}$ diagrammatically shown in Fig. 2 (b) with the vertex correction given in Fig. 2 (c). Here, both results are the same because the SCHF approximation for $\Sigma(k)$ and $\chi_{nn}$ satisfies the microscopic conservation laws.\cite{Baym} 

\begin{figure} 
\centerline{\epsfig{file=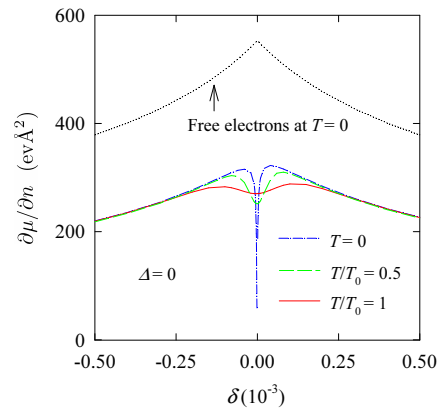,width=6. cm}}
\caption{(color online) $(\partial\mu/\partial n)_T$ as function of doping concentration $\delta$ at $T = 0$ (blue dash-dot line), $T/T_0 = 0.5$ (green dashed line), and $T/T_0 = 1$ (red solid line) for $\Delta = 0$. The black dotted line is the result of the free electrons at $T = 0$. } 
\end{figure} 

In Fig. 5, we show $(\partial\mu/\partial n)_T$ as a function of $\delta$ for $\Delta = 0$ at various temperatures. At low doping close to $\delta = 0$, there is a sharp decrease in $(\partial\mu/\partial n)_T$ at T = 0 with decreasing $\delta$. But even close to $\delta = 0$, $(\partial\mu/\partial n)_T$ is positive. This result is different from the existing calculation based on the perturbative HF approach on a bilayer graphene model.\cite{Kusminskiy2} With increasing the temperature, $(\partial\mu/\partial n)_T$ at zero doping increases and the function at low doping tends to show a flattened behavior. Since $(\partial\mu/\partial n)_T > 0$, the homogeneous electron system of $\Delta = 0$ is mechanically stable at any doping and any temperature. The result for the free electrons at $T = 0$ is also plotted in Fig. 5, which is a monotonically decreasing function of $|\delta|$. Recent experiments on the compressibility \cite{Henriksen,Young} seem qualitatively consistent with this free electron behavior. Why the Coulomb effect is not observed is still an open question.

Shown in Fig. 6 is $(\partial\mu/\partial n)_T$ as a function of $\delta$ for $\Delta/E_0 = 0.02$ at various temperatures. At T = 0, $(\partial\mu/\partial n)_T$ is negative at small $\delta$ consistent with the negative slope of $\mu$ as shown in Fig. 4. This means that the homogeneous electron system with $(\partial\mu/\partial n)_T < 0$ is unstable and favors to become an inhomogeneous state. At finite temperature, in principle, $(\partial\mu/\partial n)_T$ is positive at $\delta = 0$ because of the behavior of $\mu$ as shown in Fig. 4. At $T > 0$, it is seen from Fig. 6 that $(\partial\mu/\partial n)_T$ at $\delta = 0$ is likely infinitive. At finite but low temperature, $(\partial\mu/\partial n)_T$ decreases within a small region close to $\delta = 0$, then increases, and finally decreases slowly at large $\delta$. As shown in Fig. 6, at $T/T_0 = 1$, $(\partial\mu/\partial n)_T$ is positive at all the doping concentrations. We then deduce that the homogeneous system is stable at any doping concentrations for $T/T_0 > 1$.

\begin{figure} 
\centerline{\epsfig{file=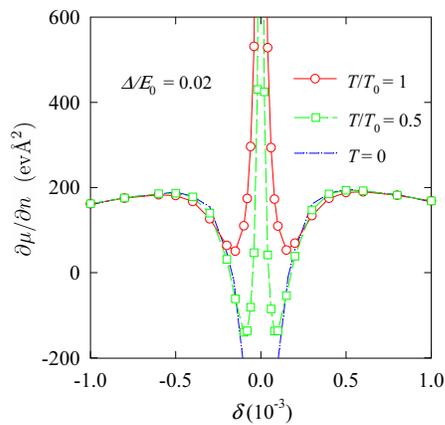,width=6. cm}}
\caption{(color online) $(\partial\mu/\partial n)_T$ as function of doping concentration $\delta$ for gapped system of $\Delta/E_0 = 0.02$ at $T = 0$ (blue dash-dot line), $T/T_0 = 0.5$ (green dashed line with squares), and $T/T_0 = 1$ (red line with circles). } 
\end{figure} 

To know whether the system is stable or not is helpful for both the theoretical understanding of the thermodynamic properties and the technological application of bilayer graphene. We have determined the phase diagram of the electron system with $\Delta/E_0 = 0.02$ as shown in Fig. 7. The red curve in Fig. 7 divides the $T-\delta$ plane into homogeneous stable ($\kappa > 0$) and unstable ($\kappa < 0$) regions. Note that at the border, $\kappa$ and $\chi$ go to infinity. $\kappa \to \infty$ means the system undergoes a phase separation or Wigner crystallization. $\chi \to \infty$ implies there is ferromagnetization in the system. Therefore, in the region with $(\partial\mu/\partial n)_T < 0$, the case of an inhomogeneous system, a Wigner crystal,\cite{Dahal} or a ferromagnetic state \cite{Nilsson1} may be possible.

\begin{figure} 
\centerline{\epsfig{file=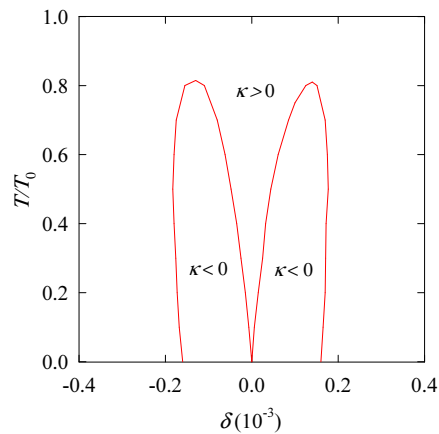,width=6. cm}}
\caption{(color online) $T-\delta$ phase diagram of the typical gapped electron system of $\Delta/E_0 = 0.02$ in bilayer graphene. The red curve is the border between the homogeneous stable ($\kappa > 0$) and the unstable ($\kappa < 0$) states.} 
\end{figure} 

In summary, we have investigated the compressibility instability of the interacting electrons in bilayer graphene using the self-consistent Hartree-Fock approximation. We have studied the combined effects due to the Coulomb interactions and the energy gap between the valence and conduction bands in the chemical potential and the compressibility of the electrons. We find that the homogeneous system with zero gap is always stable. For a gapped system, the compressibility becomes negative at low carrier doping concentrations and low temperature, leading to the instability of the homogeneous system. The phase diagram distinguishing the stable and unstable regions of a typical gapped homogeneous system is given by the present calculation.

This work was supported by the Robert A. Welch Foundation under Grant No. E-1146, the TCSUH, the National Basic Research 973 Program of China under Grant No. 2011CB932700, NSFC under Grants No. 10774171 and No. 10834011, and financial support from the Chinese Academy of Sciences for advanced research.

\end{document}